\documentclass[8.5pt,twoside,twocolumn]{article}
\usepackage[super,sort&compress,comma]{natbib} 
\usepackage{times}
\usepackage{sectsty}
\usepackage{balance} 
\usepackage{graphicx} %eps figures can be used instead
\usepackage{lastpage}
\usepackage[format=plain,singlelinecheck=false,font=small,labelfont=bf,labelsep=space]{caption} 
\usepackage{fancyhdr}
\usepackage{float}       % For creating charts, graphs and schemes

\usepackage{amssymb,amsmath}

\begin{document}

%%%%%%%%%%%%%%%%%%%%%%%%%%%%%%%%%%%%%%%%%%%%%%%%%%%%%%%%%%%%%%%%%%%%%%%%%%%%%%%%%%%%%%%%%%%%%%%%%%%%%%%%%%%%%%%
\twocolumn[
  \begin{@twocolumnfalse}

\noindent\LARGE{\textbf{Measuring the  configurational temperature of a binary disc packing}}
\vspace{0.6cm}

\noindent\large{\textbf{Song-Chuan Zhao  and Matthias Schr\"{o}ter$^{\ast}$}}\vspace{0.5cm}

\noindent \normalsize{
Jammed packings of granular materials differ from systems normally described by statistical mechanics
in that they are athermal.
In recent years a statistical mechanics of static granular media has emerged where
the thermodynamic temperature is replaced by a configurational temperature $X$ 
which describes how the number of mechanically stable configurations depends on the volume.
Four different methods have been suggested to measure $X$. Three of them are computed from 
properties of the Voronoi volume distribution, the fourth takes into account the 
contact number and the global volume fraction.
This paper answers two questions using experimental binary disc packings:
First we test if the four methods to measure compactivity provide identical results when applied to the same 
dataset. We find that only two of the methods agree quantitatively. Secondly, we test if $X$ is indeed an 
intensive variable; this becomes true only for samples larger than roughly 200 particles.
This result is shown to be due to recently measured correlations between the particle volumes
[Zhao \textit{et al., Europhys. Lett.}, 2012, \textbf{97}, 34004].}
\vspace{0.5cm}
 \end{@twocolumnfalse}
]

%%%%%%%%%%%%%%%%%%%%%%%%%%%%%%%%%%%%%%%%%%%%%%%%%%%%%%%%%%%%%%%%%%%%%%%%%%%%%%%%%%%%%%%%%%%%%%%%%%%%%%%%%%%%%%%

\section{Is there a well defined configurational temperature?}

\footnotetext{\textit{Max Planck Institute for Dynamics and Self-Organization (MPIDS), 37077 Goettingen, Germany}}
\footnotetext{\textit{$\ast$} matthias.schroeter@ds.mpg.de}

Temperature is the concept that helps us to understand how the exchange of energy stored in the microscopic degrees of freedom 
follows from the accompanying change of entropy of the involved systems.  
If we coarse-grain our view to the macroscopic degrees of freedom of particulate systems,  
such as foams or granular gases, we can still define effective temperatures 
that describe their dynamics \cite{ono_effective_2002,brilliantov:04}. This approach defines these systems 
as dissipative; kinetic energy of the particles is irrecoverably lost to microscopic degrees of freedom.

In the absence of permanent external driving such system will always evolve towards a complete rest.
And in the presence of boundary forces or gravity this rest state will be characterized by  
permanent contacts between the particles which allow for a mechanical equilibrium. 
Shahinpoor \cite{shahinpoor:80} and Kanatani \cite{kanatani:81} were the first to suggest that
such systems might still be amenable to a statistical mechanics treatment.
Sam Edwards and coworkers\cite{edwards_theory_1989, mehta:89}  have then developed this idea into a 
full statistical mechanics of static granular matter by using the ensemble of all 
mechanically stable states as a basis. 
A  crucial requirement of such an approach is the existence
of some type of excitation which lets the system explore 
the phase space of all allowed static configurations. 
This could e.g.~be realized by tapping, cyclic shear, or flow pulses of the interstitial liquid.
While there are promising results, the feasibility 
of this approach is still under debate \cite{picaciamarra:12}. 

A second key concept in Edwards' approach is the replacement of 
the energy phase space by a volume phase space where the volume function $W(\mathbf{q})$
takes the role of the the classical Hamiltonian. 
The configurations $\mathbf{q}$ represent the positions and orientations of all grains.
One can then define an analog to the partition function $Z(X)$ :
\begin{equation}
 Z(X) = \int e^{-W(\mathbf{q})/X}\Theta(\mathbf{q}) \mathrm{d}\mathbf{q}
\label{eq:partition_function}
\end{equation}
where $\Theta(\mathbf{q})$ limits the integral to mechanically stable configurations.
$X$ is the configurational temperature or compactivity, which is defined 
as $X=\partial V/\partial S$.  The configurational entropy $S$ corresponds to 
the logarithm of the number of mechanically stable configurations for a given set of boundary condition.

$S$ is neither known from first principles (except for model systems \cite{monasson:97,bowles:11}) 
nor can it be measured directly. Therefore "thermometers" measuring $X$ have to exploit other relationships;
four different ways to do so have been suggested. In this paper we will test all four of them using the
same dataset of mechanically stable disc packings.

 First $X$ can be determined from the steady state volume fluctuations 
using an analog to the relationship between specific heat and energy fluctuations
 ~\cite{nowak_density_1998,schroter_stationary_2005,ribiere:07,briscoe:08}; we will refer to this
compactivity as $X_{VF}$.
A second method is based on the probability ratio of overlapping volume histograms 
\cite{dean_possible_2003,mcnamara_measurement_2009}, allowing us to compute $X_{OH}$. 
A third way \cite{aste_invariant_2007,aste_emergence_2008}
computes $X_{\Gamma}$ from Gamma function fits to the volume distribution.  
Finally, it has been suggested  recently \cite{blumenfeld:12}, that an analysis 
based on so called quadrons \cite{blumenfeld:03} instead of Voronoi cells leads to an 
expression for $X_Q$ that involves the average particle volume and the contact number.

This difference in approach immediately raises the question if these four methods provide identical 
results when applied to the same experiment. There have been two previous experiments addressing this question:
McNamarra {\it et al.}~\cite{mcnamara_measurement_2009} found that  $X_{VF}$ and  $X_{OH}$ 
agree for tapped packings of approximately 2000 glass spheres. The same result has been reported by Puckett
and Daniels \cite{puckett:13} for compressed disc packings.

The full description of a granular packing has to take the boundary stresses into account 
\cite{wang:10,pugnaloni:11,blumenfeld:12}. The stress dependence of the entropy then 
gives rise to a tensorial temperature named angoricity. Angoricity has been computed from numerics
using the overlapping histogram method \cite{henkes:07} and from experiments using both 
fluctuations and overlapping histograms \cite{puckett:13}. 
As the experiments described below are performed in an open cell with gentle driving, the boundary stresses
can be assumed to be small and constant; we therefore exclude angoricity from our further analysis. 
  
%%%%%%%%%%%%%%%%%%%%%%%%%%%%%%%%%%%%%%%%%%%%%%%%%%%%%%%%%%%%%%%%%%%%%%%%%%%%%%%%%%%%%%%%%%%%%%%%%%%%%%%%%%%%%%%
\begin{figure}[t]
\centering
\includegraphics[width=0.4\textwidth]{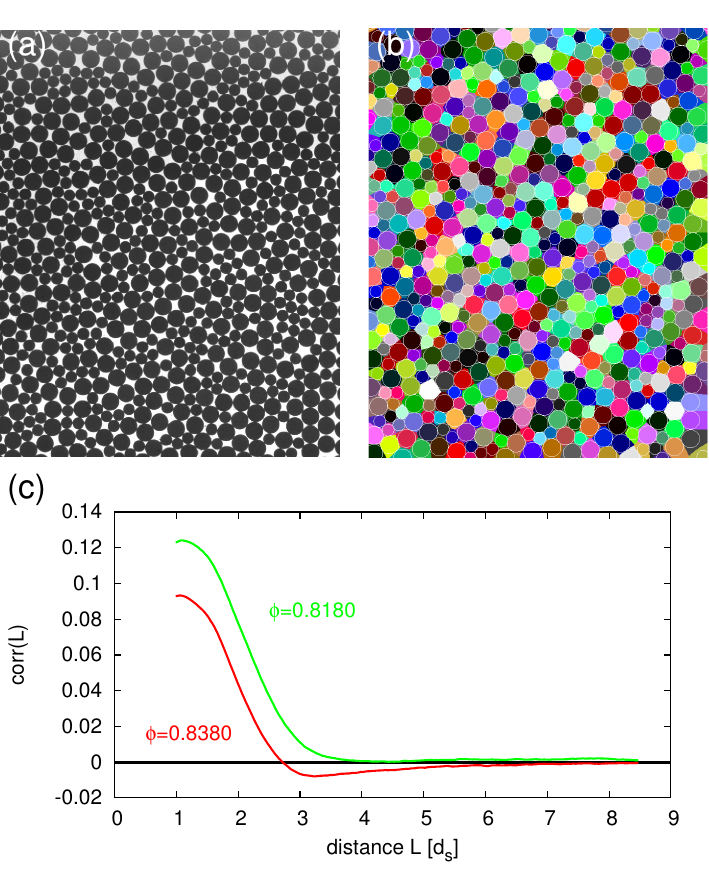}
\caption{Correlations in binary disc packings. a) experimental image and b) the corresponding Voronoi tessellation. 
c) The correlation between two Voronoi volumes as a function of their distance $L$ (measured in units of 
small disc diameters $d_s$). Dense packings exhibit anti-correlations for $L$ larger than approx. 2.5 $d_s$.  
Reproduced from  \cite{zhao_correlation_2012}.
}
\label{f.exp_corr}
\end{figure}

\section{Correlations in the Voronoi-volume of disc packings}
The three methods to compute $X_{VF}$, $X_{OH}$, and $X_{\Gamma}$ all start from the 
Voronoi-volume distribution.  However, it has been shown that the Voronoi volumes 
inside a sample are correlated \cite{lechenault:06}.  We recently  
measured the spatial extension of this correlations and demonstrated the existence of 
additional anti-correlations  between the volumes \cite{zhao_correlation_2012}. 
These correlations raise the question if $X$ is indeed an intensive parameter 
i.e.~if its value is independent of the number $N$ of particles  analyzed.
As the Voronoi-volumes reported in reference \cite{zhao_correlation_2012}
will be the basis of this study we quickly recap the relevant experimental procedures and 
results, more details can be found in the original publication. 
The disc coordinates of all configurations and volume fractions
can be downloaded from the Dryad repository \cite{dryad}.

\subsection{Experimental setup}
The experiment is performed in an air-fluidized bed
filled with a binary mixture of Teflon discs with $d_s$ = 6mm  and $d_l$ = 9 mm diameter. Data sets of 
8000 different mechanically stable configurations are prepared by repeated air pulses. Changing pressure
and duration of the air pulses allows us to control the average packing fractions $\phi$ in the ranges of 
0.818 to 0.838. After each flow pulse the discs come to a complete rest and are then imaged 
with a CCD camera (Fig.~\ref{f.exp_corr} a). After detecting the particle centers, the Voronoi
volume $V$ of each disc is determined (Fig.~\ref{f.exp_corr} b).  We then compute the free volume 
$V_f = V - V_{min}$ for each particle with $V_{min}$ being the volume the grain would occupy in a hexagonal
packing of identical discs. This step allows us to superimpose the results for small and large discs in the 
subsequent analysis.  

Our results will also depend on the packing fraction of the loosest possible packing  $\phi_{RLP}$ =0.811. 
This value is averaged over 10 packings prepared by slowly settling the discs in a manually decreased
air flow.

\subsection{Correlations in binary disc packings}
The correlation between the Voronoi volumes can be measured using:
\begin{equation}
\label{eq.t_corr}
corr(L) = \frac{\langle(V_{f,i}-\bar{V}_{f,i})(V_{f,j}-\bar{V}_{f,j})\rangle}{\sigma_f^2}
\end{equation}
Here $i$,$j$ are two points in the packing  which separated by a distance $L$.
The  free volumes at these points are $V_{f,i}$ and $V_{f,j}$.
Subtracted from them is the average free volume $\bar{V}_{f}$ 
of all particles at this point (averaged over all 8000 taps).
$\langle...\rangle$ indicates averaging over the 8000 packings  
and additionally 240 pairs of points $i,j$ within each packing. 
$\sigma_{f}^2$ is the variance of the free volumes averaged over the two points.
 
Figure \ref{f.exp_corr} c) depicts $corr(L)$ for two different packing fractions.
At low values of $\phi$ only positive correlation between Voronoi cells are found. 
Above $\phi_{AC}$ = 0.8277  anti-correlations appear for $L$ larger than approximately 2.5 small
particle diameters.  
We will show below that these (anti-) correlations control how $X$ becomes intensive. 

%%%%%%%%%%%%%%%%%%%%%%%%%%%%%%%%%%%%%%%%%%%%%%%%%%%%%%%%%%%%%%%%%%%%%%%%%%%%%%%%%%%%%%%%%%%%%%%%%%%%%%%
\section{Compactivity  $X_{VF}$ measured from volume fluctuations}
This methods starts from the assumption of a Boltzmann like probability distribution:
\begin{equation}
\label{eq.dist_density}
 p(\mathbf{q},X)=\frac{e^{-W(\mathbf{q})/X}}{Z(X)}\Theta(\mathbf{q})
\end{equation}

From equation \ref{eq.dist_density}  follows for the probability to observe a certain volume $V$  at a given
compactivity $X$:
\begin{equation}
\label{eq.vol_density}
\begin{split}
 p(V,X) &=\int \delta(W(\mathbf{q})-V) p(\mathbf{q},X)\mathrm{d}\mathbf{q}\\
 			 &=\mathcal{D}(V) \frac{e^{-V/X}}{Z(X)}
 			 \end{split}
\end{equation}
where the density of states
 $\mathcal{D}(V)=\int\delta(W(\mathbf{q})-V)\Theta(\mathbf{q})\rm{d}\mathbf{q}$ counts the number of the 
mechanical stable states available at volume $V$. 
Using equation \ref{eq.vol_density} we can determine the average volume $\bar{V}(X)$ as
\begin{equation}
\label{eq.v_bar}
 \bar{V}(X) = \int \mathcal{D}(V)\frac{e^{-V/X}}{Z(X)}V\mathrm{d}V\\
\end{equation}

Taking the derivative of equation \ref{eq.v_bar} with respect to $1/X$ shows that 
\begin{equation}
\frac{\mathrm{d} \bar{V}}{\mathrm{d} (1/X)} = - (V-\bar{V})^2 = - \sigma_{\bar{V}}^2
\end{equation}
on the other hand $\mathrm{d} \bar{V} / \mathrm{d} (1/X)$ can be rewritten as 
$- X^2  \mathrm{d} \bar{V} / \mathrm{d} X$ from which follows:
\begin{equation}
\label{eq.fluc-diss}
 X^2\frac{\mathrm{d}\bar{V}}{\mathrm{d}X}=\sigma_{\bar{V}}^2
\end{equation}

Nowak {\it et al.}~were the first to suggest that
integrating equation \ref{eq.fluc-diss} provides a way to compute $X$\cite{nowak_density_1998}: 
\begin{equation}
\label{eq.chi_var}
 \frac{1}{X_{VF}} = \int_{\bar{V}}^{\bar{V}_{RLP}}\frac{\mathrm{d}\bar{V}}{\sigma_{\bar{V}}^2}
\end{equation}
where the compactivity at the loosest possible packing has been set to infinity: $X(\phi_{RLP})=\infty$.
First measurements \cite{schroter_stationary_2005,ribiere:07} of $X_{VF}$ showed that it indeed measures
a material property as it depends on the roughness of the particles. 
It has to be pointed out that this method has the  epistemological status
of a calculation rule and not of a test of the Boltzmann assumption made in equation \ref{eq.dist_density}.
This is different for the overlapping histogram method described below.

As described in section 2 we are interested in the evolution of $X$ with the size
of the analyzed region, in the following referred to as cluster. 
Therefore we continue our study by using the normalized average volume per particle
$v= 1/N \sum V/V_g$ where the sum goes over all $N$ particles inside the cluster, $V$ is the Voronoi volume
of the individual particles  and $V_g$ the volume occupied by the particle itself. As a consequence of this
choice also $X$ is dimensionless.

\begin{figure}[t]
\centering
\includegraphics[width=0.48\textwidth]{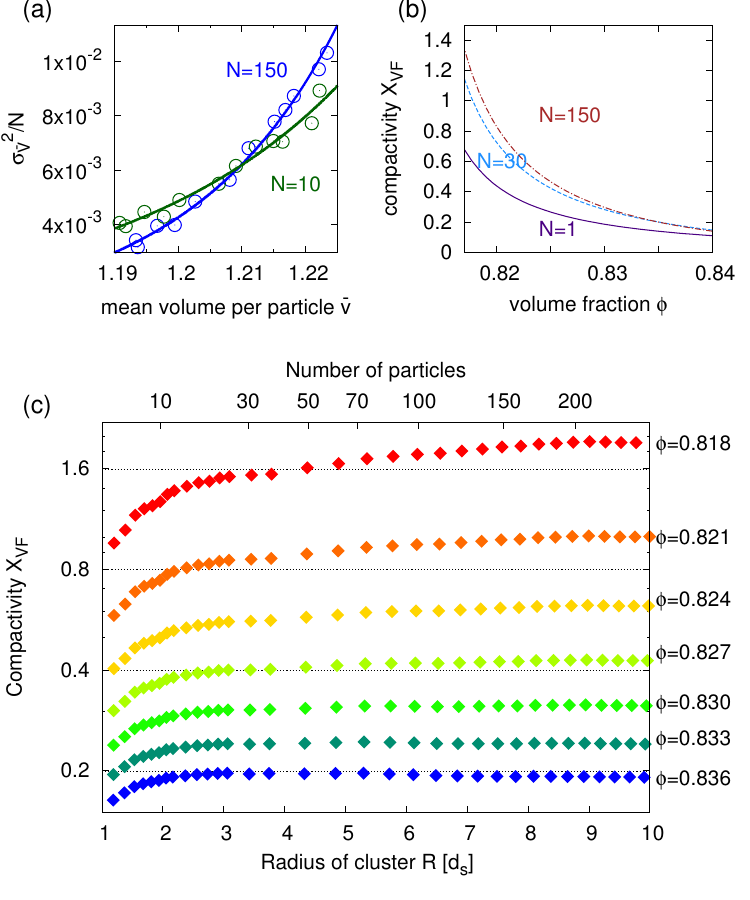}
\caption{Compactivity $X_{VF}$ measured from volume fluctuations.
a): Average volume variance $\sigma^2_{\bar{v}}$ versus average volume $\bar{v}$ measured
for clusters of size $N=10,150$. 
The solid curves are power law fits which are then used to numerically integrate equation \ref{eq.chi_var}. 
(b): The compactivity $X_{VF}$ computed from equation \ref{eq.chi_var} for three different cluster sizes. 
(c): The evolution of $X_{VF}$ with cluster sizes at different values of $\phi$. 
The radius $R$ of the analyzed cluster is proportional to $\sqrt{N/\phi}$.}
\label{f.cmp_fluc}
\end{figure}

Figure \ref{f.cmp_fluc} (a) shows that for our tapped disc packings the variance of the volume fluctuation
$\sigma^2_{\bar{v}}$ increases monotonically with the average volume $\bar{v}$ (bar meaning again the average over all 8000 taps).
To compute $X_{VF}$ we perform a  power law fit to $\sigma^2_{\bar{v}}$ and use the result to
integrate equation \ref{eq.chi_var} numerically. Figure \ref{f.cmp_fluc} (b) shows how the resulting
$X_{VF}$ depends on the packing fraction at cluster sizes $N$ of 1, 30, and 150 discs. It is obvious that
is not an intensive variable for this range of $N$.

%%%%%%%%%%%%%%%%%%%%%%%%%%%%%%%%%%%%%%%%%%%%%%%%%%%%%%%%%%%%%%%%%%%%%%%%%%%%%%%%%%%%%%%%%%%%%%%%%%%%%%%
\subsection{Correlations make $X_{VF}$ non-intensive in small systems}
For a more detailed analysis we have plotted in figure \ref{f.cmp_fluc}(c)
the dependence of $X_{VF}$ on the cluster radius $R$ measured in small 
disc diameters $d_s$. Three features of $X_{VF}$ become apparent: 
\begin{enumerate}
\item $X_{VF}$  is growing monotonously for $R < 3 d_s$ for all values of $\phi$,
\item for low to intermediate values of $\phi$,  $X_{VF}$ then reaches a plateau, and
\item for the highest packing fraction  $X_{VF}$ first decreases slightly before entering a plateau.
\end{enumerate}

All three points can be understood by considering the influence of the volume correlations shown in  
figure \ref{f.exp_corr}. Equation \ref{eq.chi_var} computes $X_{VF}$ from the average variance $\sigma^2_{\bar{v}}$
inside the cluster. This variance can be decomposed in the following way: 
\begin{equation}
\begin{split}
\sigma^2_{\bar{v}} &= \frac{1}{N} \sum_k^N\sum_m^N\langle \delta v_k \delta v_m\rangle \\
&=   \frac{1}{N} \sum_k^N (\sigma_k^2+\sum_{m\neq k}\langle \delta v_k \delta v_m\rangle)
\end{split}
\label{eq.fluc}
\end{equation}
Here $\sigma_k^2=\langle {\delta v_k}^2\rangle$ is the fluctuation of a single Voronoi cell and 
$\sum_{m\neq k}\langle \delta v_k \delta v_m\rangle := I_k$ is the volume correlation between disc $k$ and 
all other discs inside the cluster. If $k$ is in the center of the cluster, then $I_k$ 
is proportional to the area enclosed by $corr(L)$, the zero line, and $L = R$ in figure \ref{f.exp_corr}(c).

This consideration allows us to explain feature 1: while $N$ goes from 2 to approximately 25
all $I_k$ are positive and growing. Consequentially,  the variance of the cluster $\sigma^2_{\bar{v}}$ 
is larger than the sum of the variance of the individual Voronoi cells $1/N \sum \sigma_k^2$
and $X_{VF}$ grows monotonously with $N$.

The second feature stems from the fact, that once disk $k$ is more than four to five $d_s$ away from the boundary 
of the cluster,  $I_k$  becomes independent of the cluster size. As the relative importance of "boundary discs" 
decreases with cluster size, $\sigma^2_{\bar{v}}$ goes to a constant and $X_{VF}$ becomes independent of $N$.

Finally, the slight decrease in $X_{VF}$ at high packing fractions and $N$ values between approximately 30 and 150
can be attributed to the anti-correlations that appear for $\phi$ larger than 0.8277; these will decrease  $I_k$ slightly
again before it reaches its plateau value.

%%%%%%%%%%%%%%%%%%%%%%%%%%%%%%%%%%%%%%%%%%%%%%%%%%%%%%%%%%%%%%%%%%%%%%%%%%%%%%%%%%%%%%%%%%%%%%%%%%%%%%%
\section{Compactivity  $X_{OH}$ measured from overlapping histograms} 
This way to compute compactivity has been first described by Dean and Lef\`evre ~\cite{dean_possible_2003}.
It uses pairs of experiments with slightly different values of $\phi$ respectively $X_{OH}$ and computes 
then the ratio of the probabilities to observe the same local volume $V$. If the assumption of a 
Boltzmann-like probability distribution, as expressed in equation \ref{eq.vol_density}, holds, this ratio
should be exponential in $V$:
\begin{equation}
\label{eq.overlap}
\frac{p(V, X_1)}{p(V, X_2)} = \frac{\frac{\mathcal{D}(V)e^{-V/X_1}}{Z(X_1)}}{\frac{\mathcal{D}(V)e^{-V/X_2}}{Z(X_2)}} 
= 
\frac{Z(X_2)}{Z(X_1)}e^{-(\frac{1}{X_1}-\frac{1}{X_2})V}
\end{equation}
By taking the logarithm on both sides we obtain:
\begin{equation}
 \label{eq.log_overlap}
Q:= \ln\frac{p_{X_1}(V)}{p_{X_2}(V)} = \left(\frac{1}{X_2}-\frac{1}{X_1} \right) V + \ln\frac{Z_{X_2}}{Z_{X_1}}
\end{equation}
Therefore the difference between two compactivities can be computed from a line fit of $Q$ versus $V$ as it was first 
demonstrated in  \cite{mcnamara_measurement_2009}.

\begin{figure}[t]
\centering
\includegraphics[width=0.49\textwidth]{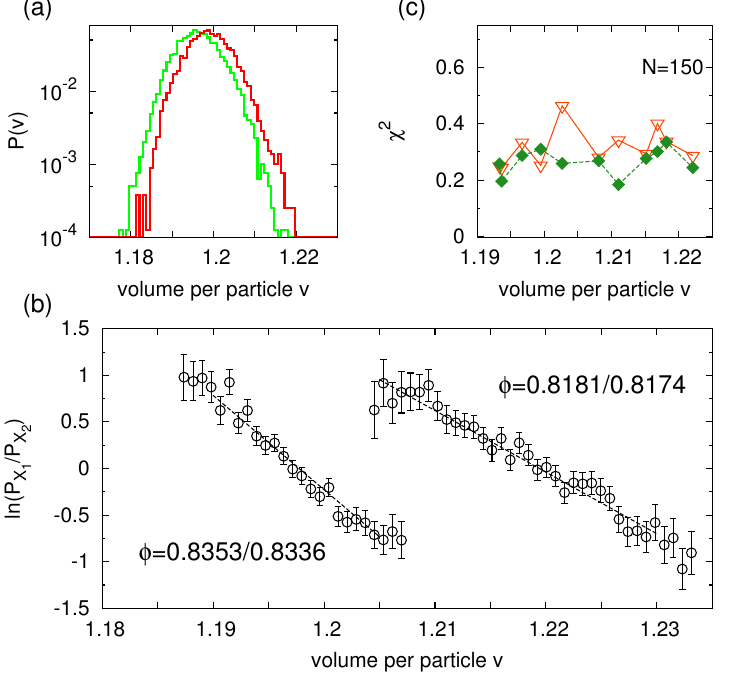}
\caption{Compactivity measured with the overlapping histogram method.  
(a) The probability to observe an average volume per particle in clusters with 150 disks. The average packing fraction 
corresponds to 0.8336 for the red curve, and 0.8353 for the green curve.
(b) The logarithm of the probabilities to observe a given volume at two different compactivities
respectively packing fractions is a linear function of the volume, which is in accordance with equation \ref{eq.log_overlap}.
Cluster size is 150 discs, the dashed lines are linear fits.
(c) The $\chi^2$ values provide a goodness of fit test for both linear (green diamonds) and parabolic (red triangle)  fits to the
probability ratios shown in panel b). The average  $\chi^2$ is 13\% smaller for the linear fit, indicating that 
a Boltzmann-like distribution is a better assumption than a Gaussian. 
}
\label{f.cmp_oh}
\end{figure}

Figure \ref{f.cmp_oh}(a) shows the distribution of average volumes for two experiments with a 
packing fraction difference of 0.0017.   
Figure \ref{f.cmp_oh}(b) demonstrates that the ratio $Q$ is indeed 
a linear function of $V$, as predicted by equation \ref{eq.log_overlap}.
By sweeping the experimentally accessible range of packing fractions, 
$X_{OH}$ can be determined from the accumulated compactivity differences
up to an additive constant $X_0$. We determine $X_0$ by setting  $X_{OH}$
for the loosest experimental packing to the value of $X_{VF}$ at this 
volume fraction.

\begin{figure}[t]
\centering
\includegraphics[width=0.49\textwidth]{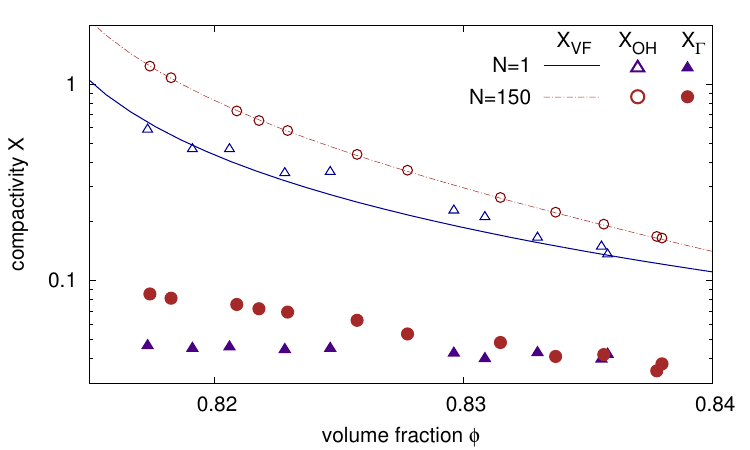}
\caption{Comparison of $X_{OH}$, $X_{VF}$, and $X_{\Gamma}$ computed for single discs and clusters of 150 particles.
In all three cases the compactivity of an individual particle is smaller than that of a larger cluster.  
$X_{OH}$ and $X_{VF}$ agree quantitatively, $X_{\Gamma}$ is about an order of magnitude smaller. 
}
\label{fig:comparison}
\end{figure}

The resulting $X_{OH}$  is shown in figure \ref{fig:comparison}. The good quantitative agreement
of $X_{OH}$ and $X_{VF}$ is not too surprising given that a) both methods are derived from
the same probability distribution (equation \ref{eq.vol_density}) and b) our determination of $X_0$.
However, the  $X_{OH}$ method provides an additional test of the assumptions  leading to equation \ref{eq.vol_density} as we
can compare the quality of a linear fit to $Q(V)$ with fit functions derived from alternative
probability distribution functions  \cite{mcnamara_measurement_2009}. A generic candidate would be
a Gaussian distribution which results in parabolic fit. Figure \ref{f.cmp_oh}(c) demonstrates
however that a linear fit is superior, adding credibility to a Boltzmann-like approach. 
Also note that canceling the density of states in equation \ref{eq.overlap} implicitly assumes
a weak form of ergodicity; if the system explores its phase space differently at the two values of $X$ we can't 
eliminate $\mathcal{D}(V)$.

%%%%%%%%%%%%%%%%%%%%%%%%%%%%%%%%%%%%%%%%%%%%%%%%%%%%%%%%%%%%%%%%%%%%%%%%%%%%%%%%%%%%%%%%%%%%%%%%%%%%%%%%%%%
\section{Compactivity $X_{\Gamma}$ measured from a Gamma distribution fit}
This third method to compute compactivity has been suggested by Aste and Di Matteo \cite{aste_emergence_2008,aste_structural_2008};
it differs from the previous approaches that it explicitly determines the density of states $\mathcal{D}(V)$. 
Based on  the observation \cite{aste_invariant_2007} that experimental Voronoi volume distributions can be well
fit by $\Gamma$ distributions, they propose to replace equation \ref{eq.vol_density} with a 
rescaled $k\text{-Gamma}$ distribution
\begin{equation}
p(V_f)=\left(\dfrac{k}{\bar{V_f}}\right)^k\dfrac{V_f^{(k-1)}}{\Gamma(k)}\exp(-k\frac{V_f}{\bar{V_f}})
\label{eq.gamma_dist}
\end{equation}
where $\bar{V_f}$ is the mean free volume (as defined in section 2), and $k$ is the shape factor. 
They then identify 
\begin{equation}
X_{\Gamma}=\frac{\bar{V_f}}{k}
\label{eq.X_gamma}
\end{equation}
or by making use of the fact that variance $\sigma_{\bar{V_f}}^2$ of the $\Gamma$ function 
is given by:  $\sigma_{\bar{V_f}}^2 = \bar{V}^2_f / k$ they derive 
\begin{equation}
X_{\Gamma}=\frac{\sigma_{\bar{V}_f}^2}{\bar{V}_f}
\label{eq.cmp_gamma}
\end{equation}

By comparing equation  \ref{eq.vol_density} and equation \ref{eq.gamma_dist}
we can also identify the density of states: 
\begin{equation}
 \frac{\mathcal{D}(V)}{Z(X)}  =  X_{\Gamma}^{-k}\dfrac{V_f^{(k-1)}}{\Gamma(k)}
\label{eq.gamma_dos}
\end{equation}

\begin{figure}[t]
\centering
\includegraphics[width=0.49\textwidth]{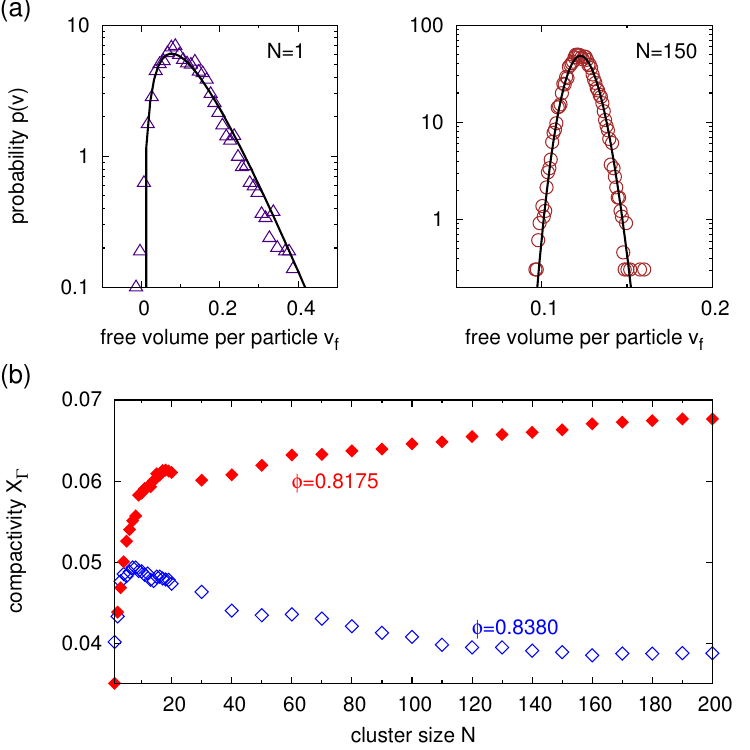}
\caption{Applying the $\Gamma$ distribution method.
(a) $\Gamma$ distribution fits to the volume distributions for a single discs and a 150 particle clusters
($\phi$ = 0.8175). 
(b) Evolution of $X_{\Gamma}$ with cluster size. 
}
\label{fig:x_gamma}
\end{figure}

Figure \ref{fig:x_gamma} (a) demonstrates that our volume distribution is well fit by a $\Gamma$ distribution. 
We then determine $X_{\Gamma}$ without any additive constant using equation \ref{eq.cmp_gamma}. 
Figure \ref{fig:comparison} show $X_{\Gamma}$ in comparison to $X_{VF}$ and $X_{OH}$. 
While all three compactivities decrease monotonously with $\phi$, the absolute values and the slope
of $X_{\Gamma}$ are quite different from $X_{VF}$ and $X_{OH}$.

Figure \ref{fig:x_gamma} (b) shows the evolution of $X_{\Gamma}$ with cluster size.
A comparison with $X_{VF}$, displayed 
figure \ref{f.cmp_fluc} (c), shows a qualitative similar influence of the volume correlations:
while $X_{\Gamma}$  increases towards a plateau
for small values of phi, it  goes through a maximum before reaching a smaller plateau value for the densest packings.

%%%%%%%%%%%%%%%%%%%%%%%%%%%%%%%%%%%%%%%%%%%%%%%%%%%%%%%%%%%%%%%%%%%%%%%%%%%%%%%%%%%%%%%

\section{Compactivity $X_{Q}$ measured from quadron tessellation}
In a  recent paper Blumenfeld {\it et al.}~\cite{blumenfeld:12}  presented an 
analysis of the statistical mechanics of two-dimensional packings based on  quadrons \cite{blumenfeld:03} 
as the building blocks of the tessellation. An advantage of this choice is that the  quadrons 
take by design into account the structural degrees of freedom of the individual particles. 
A drawback is that quadrons are not necessarily volume conserving in the presence of 
non-convex voids formed by "rattlers", i.e.~particles lying at the bottom of larger voids.
\cite{picaciamarra:07,blumenfeld:07}.
Blumenfeld and coworkers then derive the partition functions of the volume ensemble $Z_V$, 
the force ensemble $Z_F$, and the total
partition function $Z$, showing in this process that $Z \ne Z_v Z_F$. 
Finally, they obtain from $Z$ 
an expression for $X_Q$ of an isostatic packing:
\begin{equation} 
X_Q = \frac{2}{\bar{z} N + 2 M} \langle V \rangle
\label{eq:x_q_both}
\end{equation} 
$\langle V \rangle$ is the average volume of a cluster, {$\bar{z}$ the average number of contact a particle has, and
$M$ is the number of boundary forces.

If $X_Q^V$ is derived from $Z_V$ instead of $Z$ one obtains:
\begin{equation} 
X_Q^V = \frac{2}{\bar{z} N} \langle V \rangle
\label{eq:x_q_v}
\end{equation} 

\begin{figure}[t]
\centering
\includegraphics[width=0.49\textwidth]{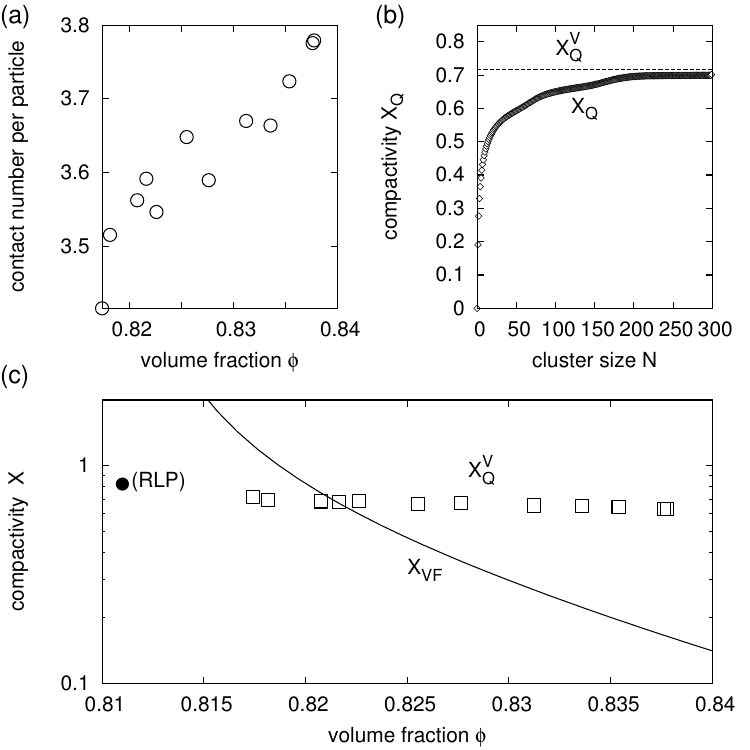}
\caption{Compactivity measured with the quadron tessellation method.
(a) In all our experiments the contact number per particle is above the isostatic value of 3.
(b) A comparison of $X_Q$ computed from the full partition function (eq.~\ref{eq:x_q_both}) 
and  $X_Q^V$ derived  from the volume ensemble (eq.~\ref{eq:x_q_v}) only. Both are computed for  
$\phi = $   0.8175. The difference vanishes in the large system limit.
(c) A comparison of $X_{VF}$ (measured for $N = 150$) and $X_Q^V$. 
The filled circle is computed for 
our Random Loose Packing value which is presumably the only isostatic point in our dataset. 
The open squares assume that  equation \ref{eq:x_q_both} is also valid for hyperstatic packings.
}
\label{fig:quadron}
\end{figure}

In comparing our results with this approach we have to acknowledge two differences. First, our experimental packings
are clearly subject to volume forces due to gravity. And secondly they are, as shown in figure \ref{fig:quadron} (a), hyperstatic,   
i.e.~their contact number is larger than what is required to fix all their mechanical degrees of freedom.  

Generally, frictional particles only become isostatic at Random Loose Packing and zero pressure \cite{shundyak:07,henkes:10}. 
Therefore the only direct comparison possible is at our RLP value $\phi$ = 0.811, the corresponding $X_Q$ is shown as 
black circle in figure \ref{fig:quadron} (c). If we assume that we can replace $\bar{z}$ in equation \ref{eq:x_q_both}
with $\bar{z}(\phi)$ and use the contact numbers displayed in figure \ref{fig:quadron} (a), we can also compute 
$X_Q$ for a larger range of $\phi$. These results are indicated as the open squares in figure \ref{fig:quadron} (c).

As $X_Q$ is only computed from the average volume of a cluster, it is insensitive to the correlations described in section 2.
On the other hand there exists a finite size effect if $X_Q^V$ is computed from the volume ensemble, 
ignoring the boundary forces. Figure \ref{fig:quadron} (b) shows how the difference between the compactivities computed 
from the full and the volume ensemble vanishes with increasing cluster size and consequentially decreasing contribution of $M$.

%%%%%%%%%%%%%%%%%%%%%%%%%%%%%%%%%%%%%%%%%%%%%%%%%%%%%%%%%%%%%%%%%%%%%%%%%%%%%%%%%%%%%%%%%%%%%%%%%%%%%%
\section{Which is the correct way to measure compactivity?}
Figures \ref{fig:comparison} and \ref{fig:quadron} (c) show clearly that the four different methods
to compute compactivity do not agree quantitatively. While $X_{OH}$ and $X_{VF}$ are identical 
within experimental scatter, $X_{\Gamma}$ is more 
than one order of magnitude smaller than $X_{OH}$ and $X_{VF}$ and at the same time decreasing less steeply  with 
volume fraction\cite{note:1}. The values of $X_Q$ are closer to  $X_{OH}$ and $X_{VF}$ again, but the evolution with $\phi$ is similar to
 $X_{\Gamma}$.  

Our data can be used to self-consistently compute all four different versions of compactivity, 
consequentially they are unsuitable as a basis to judge the correctness of the different approaches.
A potential experimental or numerical test would need an independent determination of the compactivity
of a model system from the knowledge of its entropy as a function of volume. Alternatively, if
the granular statistical mechanics could be advanced to make predictions of granular behavior 
(e.g.~segregation \cite{arxiv:12}) based on specific values of $X$, these could be tested versus
the four different "thermometers".

However, we can elucidate the difference between the four methods by computing the density of states and
comparing it to what is known to about RLP, the loosest possible, mechanically stable packing.

\subsection{Computing the density of states}
A probability distribution of the type of equation \ref{eq.dist_density} is a generic feature 
for every statistic theory that maximizes the amount of entropy represented by $p$, given certain
constraints \cite{jaynes_information_1957}. In this spirit the density of states $\mathcal{D}(V)$ 
in equation \ref{eq.vol_density} can be interpreted as encapsulating the physics of the specific 
system under consideration, while the exponential term represents our lack of knowledge about the microscopic 
state. From this perspective the main difference between the four methods is the way they determine
 $\mathcal{D}(V)$: while $X_{VF}$ and $X_{OH}$ treat it as an experimental input parameter,  
$X_{\Gamma}$ and $X_{Q}$ do provide predictions for its dependence on $V$. In this subsection 
we will however not discuss $X_Q$ as the theory has not been expanded yet to hyperstatic packings.

 $\mathcal{D}(V)$ can be computed following McNamara {\it et al }\cite{mcnamara_measurement_2009}.
Starting by rewriting equation \ref{eq.vol_density} to
\begin{equation}
\mathcal{D}(V) = p(V,X) \,  e^{V/X} \, Z(X) ,
\end{equation}
we remove the dependence on the unknown partition function by taking the ratio with $\mathcal{D}$
at a given volume $V_0$:
\begin{equation}
\label{eq:dos}
 \frac{\mathcal{D}(V)}{\mathcal{D}(V_0)} = \frac{p(V,X)}{p(V_0,X)}   e^{1/X(V-V_0)}
\end{equation}

The results are presented in figure \ref{fig:DOS} where we have chosen $V_0$ = 1.2.
Panel (a) shows that the density of states ratios computed from all eleven values of $X_{OH}$ overlap, as it
can be expected if the different preparation protocols sample the phase space with the same probabilities. 
Figure \ref{fig:DOS} (b) shows that
this agreement of the density of states is not obtained if the ratio is 
computed from $X_{\Gamma}$. 

Figure \ref{fig:DOS} provides also some insight how the configurational entropy $S$ depends on on the volume:
\begin{equation}
S = k_E \ln \left( \mathcal{D}(V) \delta V \right) = k_E \ln \mathcal{D}(V) + k_E \ln \delta V
\label{eq:entropy} 
\end{equation}
where $k_E$ is the equivalent to the Boltzmann constant. As our system is large enough and 
not constrained by boundaries (for a counter example see \cite{gao:06}), we obtain a smooth  
function $\mathcal{D}(V)$ and can therefore choose the integration interval $\delta V$ 
small enough so that its contribution to $S$ vanishes.

\begin{figure}[h]
\centering
\includegraphics[width=0.49\textwidth]{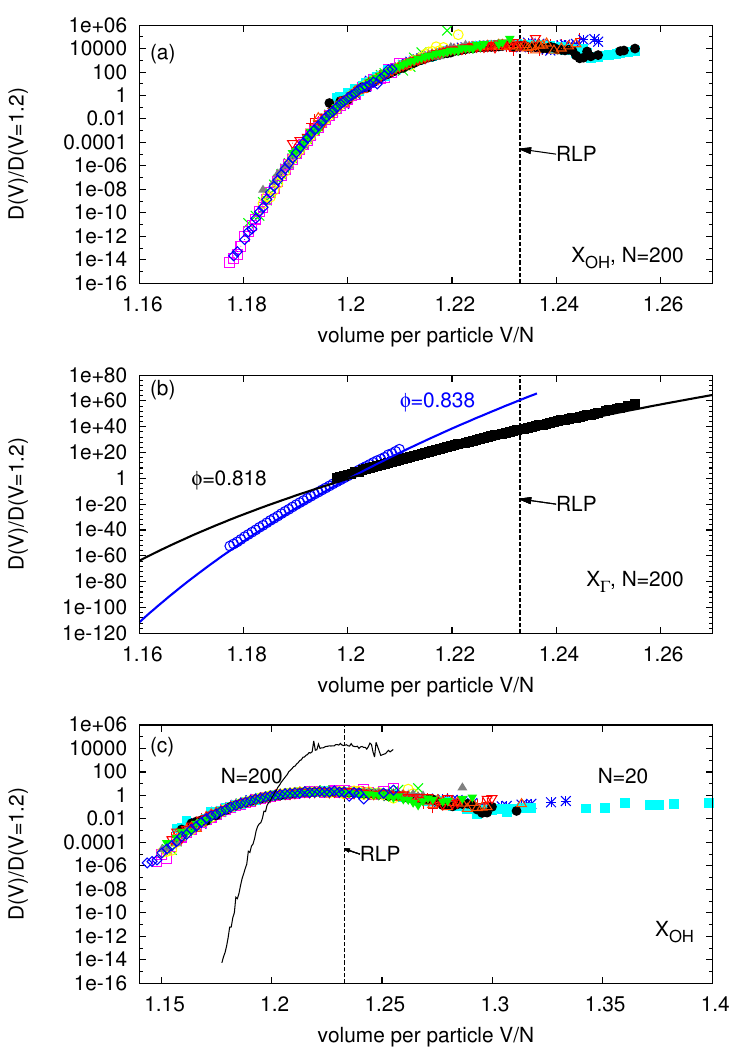}
\caption{The density of states ratio.
Panel {\bf (a)} is based on the eleven measurements of $X_{OH}$ for cluster size $N$ = 200.
The dashed line indicates to the global Random Loose Packing volume.
Panel {\bf (b)} is computed from  $X_{\Gamma}$ of the densest and loosest
packings. Points represent the measured distributions, solid lines 
are analytical results based on the Gamma functions (eq.~\ref{eq.gamma_dos}). 
In contrast to panel (a) the two curves do not overlap.
{\bf(c)} The density of states does depend on the cluster size. The 
solid, black curve corresponds to the average of all data points with $N$ = 200 in panel (a), 
the individual data points are computed from $X_{OH}$ measured at $N$ = 20.  
}
\label{fig:DOS}
\end{figure}

\subsection{Random Loose Packing}
Random Loose Packing (RLP) is first of all defined  phenomenologically as the loosest possible 
packing which is still mechanically stable, i.e.~has a finite yield stress. Mechanical stability 
requires the packing to be at least isostatic: the average number of contacts of a particle needs to be large enough
to provide sufficient constraints to fix all its degrees of freedom \cite{shundyak:07,song:08,henkes:10}. 
Experimentally it has been found that $\phi_{RLP}$ of spheres is approximately 0.55 \cite{onoda:90,jerkins:08,farrell:10}
 with the precise value depending 
on the friction coefficient $\mu$\cite{jerkins:08,farrell:10} and the confining pressure\cite{jerkins:08}. 
For binary disc packings  $\phi_{RLP}$ does also depend on their diameter ratio; for the particles in this paper
we measured  $\phi_{RLP}$ = 0.811. 

While RLP seems to imply that there exist no packings with lower $\phi$, exactly such states have been 
identified in MD simulations with $N$ = 20 discs; the new lower boundary where such states vanish has been
named Random Very Loose Packing (RVLP) \cite{pica-ciamarra:08}.  An explanation why these states between RLP and RVLP are
usually not observed in experiment is given by their small basin of attraction in phase space.

In general, the statistical mechanics approaches to granular media\cite{edwards_theory_1989,mehta:89,picaciamarra:12}
assume the configurational temperature $X \ge 0$. Then it follows from
\begin{equation} 
\partial S = 1/X \, \partial V
\label{eq:partial_S}
\end{equation}
that the configurational entropy increases the looser the packings become.
At RLP $S$ reaches a maximum and then decays  to zero for looser configurations which by definition
are not mechanically stable packings. Further on, it is often assumed 
 that the maximum of S is analytic which implies that
$X$ becomes infinity at RLP\cite{edwards_theory_1989,mehta:89,pica-ciamarra:08}.

The density of states computed from $X_{VF}$ respectively $X_{OH}$ does agree with these predictions;
figure  \ref{fig:DOS} shows that it reaches a global maximum at RLP. 
This is however only a self-consistency test as the  value of  $\phi_{RLP}$ explicitly
entered the computation of these two compactivities. More probative is the fact that the volume fluctuations 
used to compute $X_{VF}$ have been shown to depend on $\mu$ \cite{schroter_stationary_2005}.
Consequentially, also a configurational entropy computed  from $X_{VF}$ will depend on $\mu$, which is a 
necessary condition as the number of mechanically stable configurations is also depending on $\mu$. 

Figure  \ref{fig:DOS} c)  shows that the density of states does also depend on the size of the system.
Systems with 20 discs posses considerably looser states than those with 200 discs. 
This effect points to an explanation of the configurations between RLP and RVLP as a consequence 
of the finite size of a system, they will vanish in the thermodynamic limit.

The density of states computed from $X_{\Gamma}$  (figure \ref{fig:DOS} b) seems not to display a maximum 
at $\phi_{RLP}$. However, RLP might become the most likely state if in the limit $N \rightarrow \infty$ the states 
between RLP and RVLP vanish. 
Either way,
$X_{\Gamma}(\phi_{RLP})$ will be finite and display a jump to approximately half its value 
for the  "fluid" configurations below  $\phi_{RLP}$ \cite{aste_structural_2008}, which have Voronoi volume distributions that are also
well described by Gamma functions\cite{senthil_kumar_voronoi_2005}. While this behavior might be not the canonical expectation, especially
as $X_{\Gamma}$ is undefined for fluid, non-stable configurations, it does not contradict experimental results.
This is different for the influence of friction: it has been shown that the Gamma distribution fits are 
independent of $\mu$\cite{aste_invariant_2007,aste_emergence_2008}, consequentially neither $\phi_{RLP}$, nor $X_{\Gamma}$ and the derived 
$S$ will reflect the different number of mechanical stable states resulting from changes in friction.

Finally, it is difficult to comment on the relationship between $X_Q$ and RLP as the theory is presently only worked out at exactly RLP.
$X_Q(\phi_{RLP})$ is of finite value and will depend only very weakly on $\mu$ via $V$.
If we allow similar to $X_{\Gamma}$ for its existence also for $\phi < \phi_{RLP}$, 
$X_Q$ will be depending on the preparation protocol because the contact number is protocol dependent below RLP\cite{heussinger:09}.

\subsection{Open questions}
Because granular matter is athermal, the question of ergodicity has to be answered separately for
each experimental or numerical protocol used to explore the phase space of mechanically stable configurations.
An indication of their differences is e.g.~the different $\phi$ dependence of the volume fluctuations  
when the sample is either mechanically tapped\cite{ribiere:07} or excited by flow pulses\cite{schroter_stationary_2005}.
Nonergodicity has recently also been shown for numerical tapping of frictionless hard spheres\cite{paillusson:12}, however
there the analysis was not restricted to mechanically stable states. In contrast, in numerical simulations of frictional discs
fluidized with flow pulses an equivalence between time and ensemble averages has been found\cite{pica-ciamarra:06}.

The biggest challenge to the concept of a configurational granular temperature are however 
recent experiments by Puckett and Daniels\cite{puckett:13}
where they studied the volume and force fluctuations of two samples of discs which where in mechanical contact. 
They found that the angoricity of the two samples equilibrated, but the compactivity (computed as $X_{OH}$ and $X_{VF}$)
did not. However, their preparation protocol was biaxial compression which did not allow for strong particle 
rearrangements, therefore it might have been prohibiting volume exchange between the two subsystems by quenching too fast.  
Clearly more work is needed to answer the question if compactivity is a variable predictive of granular 
behavior or just a number following from an algorithm.

\section{Conclusions}
Using the same experimental dataset, we have computed the configurational granular temperature $X$ 
with the four different methods that have been previously suggested. The two methods that treat the density
of states as an experimental input agree quantitatively with each other. The $X$ values computed from the other two methods, which specify
the density of states from independent theoretical considerations, are both different from each
other and from the first two methods. Our results do not provide a direct way to judge the correctness 
of the individual approaches. But the derived density of states shows that only the two agreeing methods
provide a satisfying explanation how Random Loose Packing depends on friction.

Our measurements also demonstrate that $X$ becomes only an intensive variable when computed for clusters of size 
$N$ larger 150 particles. This effect is due to the volume correlations of neighboring particles. Consequentially, 
individual Voronoi cells are not suitable `quasi-particles' to define a configurational temperature in granular packings. 
This will complicate the application of a statistical mechanics approach to small granular systems and in the presence
of local gradients.

We acknowledge helpful discussions with Karen Daniels and Klaus Kassner.

%\bibliographystyle{unsrt}
%\bibliography{cmp_ref}

\begin{thebibliography}{10}
\bibitem{dryad}
DOI from Dryad; data upload is only possible after paper is accepted by a
  journal.
\bibitem{note:1}
This difference can not be attributed to the use of the free volume $v_f$ for
  the computation of $X_{\Gamma}$ as $v_f$ differs from $v$ only by a constant
  $2\sqrt{3}/\pi$. Therefore neither the shape of the probability distributions
  used to compute $X_{OH}$ nor the size of the volume fluctuations, on which
  $X_{VF}$ is based on, would change if they were also computed from $v_f$.
\end{thebibliography}

\end{document}